\documentclass{aa}  

\usepackage{graphicx}
\usepackage{txfonts}
\begin{document}

   \title{The Spectrum of Pluto, 0.40 – 0.93 $\mu$m  I. Secular and longitudinal distribution of ices and complex organics}


   \author{V. Lorenzi
          \inst{1}
          \and N. Pinilla-Alonso\inst{2}
          \and J. Licandro \inst{3,4}
          \and D. P. Cruikshank\inst{5}
          \and W. M. Grundy \inst{6}
          \and R. P. Binzel\inst{7}
          \and J.P. Emery\inst{2}
          }
   \institute{Fundaci\'on Galileo Galilei-INAF, Spain\\
              \email{lorenzi@tng.iac.es}
              \and Department of Earth and Planetary Sciences, University of Tennessee, TN, USA
              \and Instituto Astrof\'isico de Canarias, IAC, Spain
              \and Departamento de Astrof\'isica, Universidad de La Laguna, Tenerife, Spain
              \and NASA Ames Research Center, CA, USA
              \and Lowell Observatory, AZ, USA
              \and Massachusetts Institute of Technology, MA, USA
             }

   \date{Received September 15, 1996; accepted March 16, 1997}

 
  \abstract
   {During the last 30 years the surface of Pluto has been characterized, 
   and its variability has been monitored, through continuous near-infrared spectroscopic observations. But in the visible range only few data are available.}
   {The aim of this work is to define the Pluto's relative reflectance in the visible range to characterize the different components of its surface, 
   and to provide ground based observations in support of the \textit{New Horizons} mission.}
   {We observed Pluto on six nights between May and July 2014, with the imager/spectrograph ACAM at the \textit{William Herschel} Telescope (La Palma, Spain). 
   The six spectra obtained cover a whole rotation of Pluto (P$_{rot}$ = 6.4 days). For all the spectra we computed the spectral slope  
   and the depth of the absorption bands of methane ice between 0.62 and 0.90 $\mu$m. 
   To search for shifts of the center of the methane bands, associated with dilution of CH$_4$ in N$_2$, we compared the bands with reflectances of pure methane ice.}
   {All the new spectra show the methane ice absorption bands between 0.62 and 0.90 $\mu$m. 
   The computation of the depth of the band at 0.62 $\mu$m in the new spectra of Pluto, and in the spectra of Makemake and Eris from the literature, 
   allowed us to estimate the Lambert coefficient at this wavelength, at a temperature of 30 K and 40 K, never measured before. 
   All the detected bands are blue shifted, with minimum shifts in correspondence with the regions where the abundance of methane is higher. 
   This could be indicative of a dilution of CH$_4$:N$_2$ more saturated in CH$_4$. 
   The longitudinal and secular variations of the parameters measured in the spectra are in accordance with results previously reported in the literature and with the 
   distribution of the dark and bright material that show the Pluto's albedo maps from \textit{New Horizons}. 
   }
   {}

   \keywords{Kuiper belt objects: individual: Pluto -- methods: observational -- methods: numerical -- techniques: spectroscopic -- planets and satellites: composition}

   \maketitle
%

\section{Introduction}

The surface of Pluto is covered with ices of varying degrees of volatility, and even at its low average temperature of $\sim$40 K, 
there is an exchange of molecules between the surface and the thin atmosphere.  Because of Pluto's high obliquity (119.6$^\circ$) and the eccentricity of its orbit 
(perihelion 29.66 AU, aphelion 48.87 AU), there are pronounced seasons over the 247.7-year orbital period.  
These factors ensure that Pluto's surface undergoes changes on a seasonal time scale. Secular changes may also occur, owing to the long-term evolution of 
the chemical components of the surface and atmosphere under the influence of solar and cosmic ray irradiation.  On the time scale of Pluto's 6.4-day rotation period, 
clear changes in the distribution of ices across its surface are seen. 

Indeed, on seasonal and rotational time scales, changes in Pluto's overall (global) brightness and in the near-infrared spectral signatures of the three principal ices, 
N$_2$, CH$_4$, and CO have been detected and quantified (\citealp{Grundy2013} (hereafter G13); \citealp{cruik2015} and references therein).
Because Pluto has only been observed over one-third of its full orbit, secular changes might be confused with seasonal variations, 
with the result that at the present time there is no unequivocal evidence for observable secular changes. 

In addition to the ices on Pluto' surface, there is another material present that imparts colors as seen in the extended visible spectral region, 0.30-1.00 $\mu$m.  
As deduced from multi-color photometric measurements and images, the colors vary from light to medium yellow, red-brown, to black \citep{Buie_b}.  
The ices detected to date do not produce these colors, and it is generally accepted that photochemical reactions in Pluto's atmosphere and in the surface ices 
create colored macromolecular matter that contains or is made of complex organic chemicals derived from carbon, nitrogen, and oxygen \citep[e.g.,][]{materese2014}. 
The coloring properties of these materials are well established \citep[e.g.,][]{khare1984,cruik2005_tholins,imanaka2012, materese2014}. 
The coloring agent is not uniform over Pluto's surface, changing noticeably over the dwarf-planet's 6.4-day rotation.

The best way to quantify the color of Pluto and to search for diagnostic characteristics is to record the reflectance properties of the surface in the photo-visual, 
or extended visual spectral region (0.30 – 1.00 $\mu$m).  
Only a few such measurements have been reported in the literature, beginning with observations made in 1970 by \cite{fix1970},  
and followed by \cite{benner1978}, \cite{bell1979} (data published in Cruikshank and Brown 1986), \cite{barker1980}, \cite{buie1987} and \cite{grundy1996}. Individually or together, 
these studies do not define a well sampled range of rotation and seasonal change, or a full sample of wavelengths in the desired range at adequate spectral resolution. 
The history of color measurements with ground-based telescopes and the \textit{Hubble Space Telescope} (HST) is reviewed by \cite{cruik2015}, 
who note that the color measured in the (B-V) photometric index was constant from 1953 to 2000 within a few tenths of a percent, but then changed abruptly in 2002.  
\cite{Buie_b} give the details of the HST observations that showed a global change in color to a redder (B-V) index that persists to the present day.

The variation of the relative amounts of ices and dark material have been mapped before (\citealt{Buie_b}; G13), 
but are even more obvious now in the light of the preliminary new images obtained by the \textit{New Horizons} mission. 
The maps combining the reflectance and colours of Pluto's surface clearly show a diverse palette from the very dark region on the equator, 
informally named \textit{Cthulhu Regio} to the ``clear peach orange'' region, informally named the \textit{Tombaugh Regio} (Stern et al., submitted). 

During 2014, we conducted a new observational study in order to provide a high-quality data set of spectrophotometric data in the 0.40 – 0.93 $\mu$m spectral range. 
These observations enable the definition of Pluto's reflectance and the investigation of weak methane ice absorption bands shortward of 1.00 $\mu$m.  
The data have hemispheric spatial resolution over a whole rotation of the Pluto's longitudes.  
The new data are compared with earlier observations to detect any changes in spectral reflectance that may have occurred in the last few decades.  
This comparison also provides a basis for the modeling of the spectrum with radiative transfer models that account for the known components of Pluto's surface as well as additional components 
that are suspected to occur.  The observations presented here include the reflected light of both Pluto and Charon. 
These observations are also in support of the \textit{New Horizons} mission 
that observed the Pluto system in July, 2015 and are part of a long term campaign to follow the secular evolution of the Pluto's surface properties.  

In the next section we describe the observations. In section 3 we analyze the obtained spectra, calculating the slopes of the relative reflectances, 
the depth and the area of the absorption bands of methane ice, and measuring the shifts of their centers. In Section 4 we discuss the results 
and in Section 5 we report our conclusions. 

\section{The observations}

We observed Pluto on six different nights between May 13 and July 17, 2014 using the low-resolution spectroscopic mode of ACAM \citep[auxiliary-port camera,][]{benn}. 
ACAM is an imager/spectrograph that is mounted permanently at a folded-Cassegrain focus of the 4.2 m \textit{William Herschel Telescope} at the Observatorio del Roque de los Muchachos (ORM), La Palma. 
We used the V400 disperser and the 1.5” slit providing a resolution of $\sim$300 at 0.55 $\mu$m (the dispersion is 3.3 \AA$/$px). 
The slit was oriented at the parallactic angle, and the tracking of the telescope was set to compensate for Pluto's proper motion.

In order to cover a whole rotation of Pluto, we originally scheduled seven observations equally spaced in time to observe Pluto's surface every $\sim$50 degrees in longitude. 
Unfortunately, we discarded one of those observations due to technical problems. The six that we kept cover 310 degrees in longitude. 
The details of the observations are described in Table \ref{tab1}. 
We define longitudes following the IAU2009 model that corresponds with the \textit{right-hand rule}. 
Every night we obtained three consecutive spectra of Pluto of 300 seconds exposure time each, wich provides a S/N > 200 in most of the wavelength range and even better after averaging the three spectra.

\begin{table*}
\caption{Details of the observations. Time = mid exposure time; am(Pluto) and am(SA) = mid exposure airmass of Pluto and the solar analogs, respectively; 
Lon = Pluto sub-Earth longitude; Lat = Pluto sub-Earth latitude; p$_v$ = Geometric albedo of Pluto at 0.56 $\mu$m.} 
\label{tab1}      
\centering                          
\begin{tabular}{l c c c c c c c}        
\hline\hline 
\textbf{2014 Date}& \textbf{Time} & \textbf{am (Pluto)} &\textbf{Solar Analog} &\textbf{am (SA)} & \textbf{Lon (deg)}&\textbf{Lat (deg)} & \textbf{p$_v$}\\    
\hline 
&&&&&&&\\
 May 13  & 04:38 & 1.5 &L112-1333 &1.6 &103.7$^\circ$ &51.2$^\circ$ &0.44  \\ 
         &       &     &L107-998  &2.3 & & &  \\        
 May 14  & 03:40 & 1.6 &L112-1333 &1.7 &49.7$^\circ$ &51.2$^\circ$ &0.48  \\
         &       &     &L107-998  &1.5 & & &  \\        
 June 6  & 01:50 & 1.7 &L112-1333 &1.5 &197.9$^\circ$ &50.9$^\circ$ &0.59  \\
         &       &     &L107-998  &1.2 & & &  \\        
 June 7  & 01:57 & 1.6 &L112-1333 &1.5 &141.3$^\circ$ &50.9$^\circ$ &0.50  \\
         &       &     &L107-998  &1.3 & & &  \\        
 June 18 & 00:23 & 1.8 &L112-1333 &2.3 &245.2$^\circ$ &50.7$^\circ$ &0.57  \\
         &       &     &P041C     &1.5 & & &  \\        
 July 17 & 22:40 & 1.7 &L112-1333 &1.7 &358.9$^\circ$ &50.0$^\circ$ &0.52  \\
         &       &     &P041C     &1.4 & & &  \\        
\hline                                   
\end{tabular}
\end{table*}

Data reduction was performed using the Image Reduction and Analysis Facility (IRAF) standard procedures \citep{tody}. Preprocessing of the CCD images included bias and flat field correction. 
Flat field images were obtained using calibration lamps (tungsten continuum lamp) in the afternoon. 
Wavelength calibration lamps (CuAr and CuNe) were obtained before and after the set of Pluto's spectra, maintaining the same position angle to avoid the effect of instrumental flexures in the wavelength calibration. 

To correct the spectra for telluric absorptions and to obtain the relative reflectance, we observed every night two different Landolt stars \citep{land} that have proven to be good solar analogs \citep{lic2}.  
We observed them at airmasses very similar to that of Pluto (see Table \ref{tab1} for the details). 
These solar analogs are bright (10 < Vmag < 12.4) so the exposure times were always shorter than 60 seconds. 
To reduce the number of bad pixels and artifacts, we obtained three spectra of each of the stars as we did for Pluto. 
The reduction of the spectra of the solar analogs was done following the same procedure as for Pluto.  
All the calibration images, the solar analogs and the Pluto images were obtained in $\sim$1 hour. 

The extinction of the sky is dependent on the wavelength, with shorter wavelengths experience greater extinction. 
To minimize the spectral extinction effect from the difference in airmass between the stars and the target, we applied color correction to the spectra of the object and the stars. 
We used the mean extinction coefficient for the Observatorio del Roque de los Muchachos \citep{king}. 

We finally divided the color-corrected spectra of Pluto by the color-corrected spectra of the stars to obtain Pluto's relative reflectance. We used the sky lines at this step to refine the wavelength calibration. 
Before averaging all of them into a ‘final’ relative reflectance, we visually inspected the 18 reflectances and discarded those that showed problems due to the systematic errors. 
We averaged the remainder of them into a final relative reflectance for each date of observation. 
 
\begin{figure}
 \begin{center}
  \includegraphics[width=9.5cm]{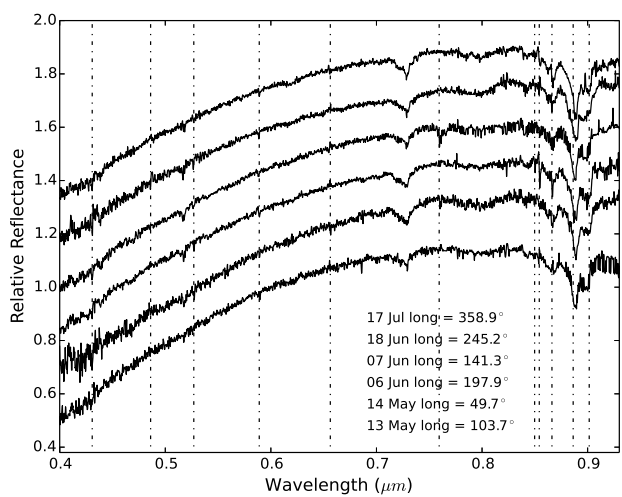}
 \end{center}
   \caption{Relative reflectances of Pluto corresponding to hemispheres centered at six different sub-Earth longitudes. 
   All the reflectances are normalized at 0.60 $\mu$m and shifted by 0.15 in relative reflectance for clarity. 
   The labels are ordered (top to bottom) in the same order than the reflectances. 
   The vertical dashed lines show the telluric absorptions. 
   Some of these telluric absorptions have not been completely removed. The absorption at 0.52 $\mu$m is an artifact, 
   it is present on the spectra of the solar analogs, its center corresponds to the magnesium triplet (0.5169 –0. 5183 $\mu$m) that is present in G stars. 
   }
   \label{fig1}
\end{figure}

In Fig. \ref{fig1} we present the six reflectance spectra obtained in this observing campaign.  These are the average of the Pluto + Charon spectra 
divided by the spectra of the solar-type stars. This division cancels the solar color and the solar and telluric absorption lines. 
The spectra have been normalized to 1.0 at 0.75 $\mu$m, to enhance the differences in the slope of the continuum below 0.70 $\mu$m.  The contribution of Charon has not been removed in this plot.  
Nevertheless, the spurious light of Charon is not a problem as Charon is two magnitudes fainter than Pluto, and its color is neutral at these wavelengths (as we explain in detail later).

\begin{figure}
 \begin{center}
  \includegraphics[width=9.5cm]{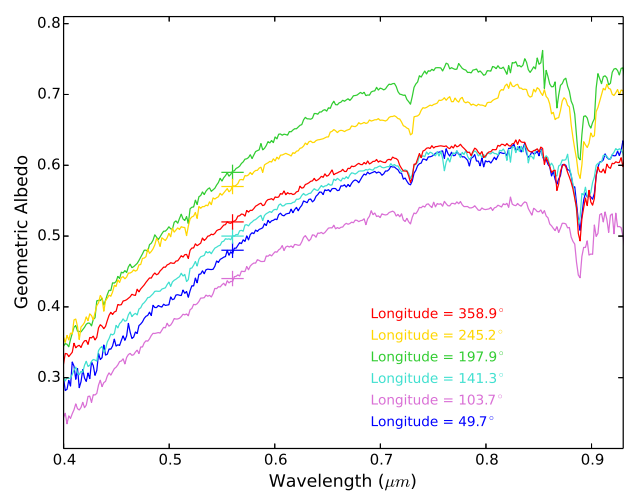}
 \end{center}
   \caption{Same spectra of Fig. \ref{fig1} smoothed using an average filter of 5 pixels and represented in geometric albedo. 
   The symbol plus indicates, for each spectrum, the global geometric albedo of Pluto alone at 0.56 $\mu$m calculated for the corresponding date of the observation.}
   \label{fig2}
\end{figure}

In Fig. \ref{fig2}, we present the data from Fig. \ref{fig1} smoothed using an average filter of 5 pixels and plotted in geometric albedo. 
To transform from reflectance to albedo, we used the magnitudes of Pluto alone from Fig. \ref{fig1} of \cite{Buie_a}, in which they plot the lightcurve of Pluto resolved from Charon.  
The values in the figure are shown as mean opposition magnitudes corrected to phase angle $\alpha = 1^\circ$.  
We have taken the \cite{Buie_a} magnitudes at $\alpha = 1^\circ$ for the longitudes we observed and have calculated the geometric albedos corresponding to those longitudes. 
Using a radius of 1184 km for the dwarf planet \citep{lellouch2014}, 
we calculated the global geometric albedo of Pluto alone at the V wavelength (0.56 $\mu$m) for each date we observed (listed in Table \ref{tab1}) using the equation
\\

-2.5 log p$_V$ = V$_{Sun}$ – V$_{Pluto}$ -5 log r + 5 log (R$\times\Delta$),
\\

\noindent
where V$_Sun$ = -26.75, r = Pluto's radius (in AU), R = Pluto's mean heliocentric distance (39.482 AU) and  $\Delta$= Pluto's geocentric distance (38.482 AU).  
A small revision of Pluto's radius from the \textit{New Horizons} investigation makes a negligible change to the albedo calculations.

We made no correction for the contribution of Charon to the color of  (Pluto + Charon) that we observed, assuming that Charon's color is neutral gray. 
From observations of the mutual transit and occultation events of the mid-1980s, \cite{binzel1988} found a nearly neutral gray color for Charon, with B–V = 0.700 $\pm$ 0.010 ((B–V)$_Sun$ = 0.65).  
More recently, \cite{Buie_a} derived (B–V) = 0.7315 $\pm$ 0.0013 using HST data.  
At the photometric B band (0.44 $\mu$m), \cite{tholen1990} give the geometric albedo of Pluto as 0.44–0.61, and that for Charon 0.38. 

\section{Results and analysis}
\label{res_an}

For each spectra we computed four parameters: 
(1) the spectral slope in order to quantify the color of Pluto at different longitudes and, thus, search for possible variations in the amount of solid complex organics over its surface; 
(2) the integrated area of the absorption bands of methane ice at 0.62 $\mu$m in order to determine if this weak CH$_4$ band is detected as in other trans-Neptunian objects (TNOs) \citep{licmak,Alvaro};  
(3) the depth of the methane ice bands 
between 0.62 and 0.90 $\mu$m in order to search for possible variations of the abundance of CH$_4$-ice on the surface of Pluto with longitude; 
(4) the shift of the center of the absorption bands of methane ice at 0.73, 0.79, 0.80, 0.84, 0.87 and 0.89 $\mu$m 
in order to search for possible variations of the dilution of CH$_4$ in N$_2$ with longitude \citep[previously reported by][]{quirico}. 

\subsection{Slope}

The slope in the visible is an indicator of the amount, particle size, or type of complex organics on the surface. 
Complex organics have been detected on the surface of asteroids (24) Themis and (65) Cybele \citep{rivkin2010,lic2011} 
and the saturnian moons Iapetus and Phoebe \citep{cruik2008, cruik2014}, 
and their presence has been inferred for other solar system objects \citep[e.g. TNOs and Titan,][]{cruik2005_tholins}. 
The recent discovery of a suite of 16 organic compounds, comprising numerous carbon and nitrogen-rich molecules, on the surface of the comet 67P/Churyomov-Gerasimenko 
\citep{wright2015,goesmann2015} 
suggests that organic compounds that are the precursor of the more complex organics (such as tholins) are a ubiquitous ingredient of the minor bodies in the Solar System. 
Not surprisingly, red slopes, typically associated with the complex organics, are common among the spectra of TNOs. 

Here we compute the slope of Pluto's visible spectrum in two ranges: from 0.40 to 0.70 $\mu$m and from 0.57 to 0.70 $\mu$m. The first range comprises the full spectral region below 0.70 $\mu$m covered by our spectra. 
The second range enables the comparison with previous spectra of Pluto, Eris, and Makemake in the literature. 
To compute the slope values in both cases, we first normalized all the spectra at 0.60 $\mu$m, and then we fitted each of them in the chosen range with a linear function. 
We made this computation for all the six spectra and for the average relative reflectance. 
The slope is expressed here as $\%/1000$\ \AA. 

To evaluate the errors in the computation of the slope, we took into account both the error in the fit and the systematic errors in the reflectance. The systematic errors depend on different contributions: e.g., errors introduced by color corrections or by dividing by the spectra of the solar analogs.  
The error introduced by dividing by the spectra of the solar analog is the dominant part of the systematic errors. To estimate this error we compared the six individual spectra of the stars observed each night following the procedure described in \cite{lor_var} and Pinilla-Alonso (submitted). 
After discarding those spectra with an extreme behavior, we took the standard deviation of the slope of these spectra as the error in the slope for the night. These values are shown in Table \ref{tab2}. We take the maximum value in the run as the systematic error. 
We found that the formal error coming directly from the computation of the slope is an order of magnitude lower than the systematic error, so we took 0.8 $\%/1000$\ \AA\ as the final uncertainty associated to the slope calculation for all the spectra. 

\begin{table}
 \caption{Estimated systematic error in the slope expressed in $\%/1000$\ \AA.}
 \label{tab2}
 \centering                          
\begin{tabular}{l c |l c}        
\hline\hline 
\textbf{2014 Date}& \textbf{Syst. error}&\textbf{2014 Date}& \textbf{Syst. error}\\
\hline
 May 13 & 0.45 & June 7  & 0.76 \\
 May 14 & 0.70 & June 18 & 0.73 \\
 June 6 & 0.37 & July 17 & 0.71 \\
\hline                                   
\end{tabular}
\end{table}

To compare with other spectra in the literature, we computed the slope of those spectra in the same way that we did for the new data presented herein. 
We used for this comparison the spectra of Pluto from \cite{grundy1996} and the spectra of two other large TNOs, (136472) Makemake from Licandro et al. (2006a) and (136199) Eris from \cite{Alvaro}.  
For the spectra from the literature of Pluto, Makemake, and Eris, there are no estimates for the systematic errors in the original papers, 
so we take the typical value of 1 $\%/1000$\ \AA\ \citep[e.g.][]{lazzaro} for all the spectra (for these spectra the systematic errors are also an order of magnitude greater than those associated with the fit calculation).  
The results are listed in Table \ref{tab3}.

\begin{table*}
 \caption{Computed spectral slope between 0.57 and 0.70 $\mu$m, and between 0.40 and 0.70 $\mu$m, 
 with the spectra normalized at 0.60 $\mu$m, for the spectra of Pluto \citep[this work;][]{buie1987, grundy1996}, 
 for the spectrum of Makemake \citep{licmak} and for the spectrum of Eris \citep{Alvaro}.}
 \label{tab3}
 \centering                          
\begin{tabular}{l c c c}        
\hline\hline 
\textbf{Longitude (degrees)}& \textbf{Slope ($\%/1000$\ \AA) 0.57-0.70 $\mu$m}&\textbf{Slope ($\%/1000$\ \AA) 0.40-0.70 $\mu$m}& \textbf{Reference}\\
\hline\\
 49.7$^{\circ}$ &14.2 &19.9 & This work \\
 103.7$^{\circ}$&12.6 &20.6 & ''\\
 141.3$^{\circ}$&12.5 &19.3 & ''\\
 197.9$^{\circ}$&12.7 &19.4 & ''\\
 245.2$^{\circ}$&10.7 &17.3 & ''\\
 358.9$^{\circ}$&10.8 &17.2 & ''\\
 Average spectrum &12.3 &19.0 & ''\\
\hline
&&&\\
 25$^{\circ}$ (1983) &  5.7&- & \cite{buie1987}\\
 329$^{\circ}$ (1983)& 5.1 &- & ''\\
 144$^{\circ}$ (1990)& 7.7 &- & \cite{grundy1996} \\
 56$^{\circ}$ (1993) & 10.8&- & ''\\
 150$^{\circ}$ (1994)&10.4 & -& '' \\
\hline
&&&\\
 Makemake&12.9 &17.3 & \cite{licmak}\\
 Eris&4.4 &6.5 & \cite{Alvaro}\\
\hline                                   
\end{tabular}
\end{table*}

\subsection{Methane absorption bands}

\begin{figure}
  \begin{center}
  \includegraphics[width=9.5cm]{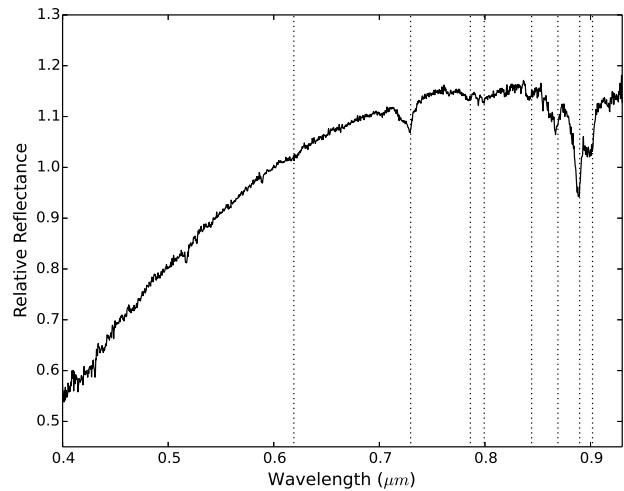}
 \end{center}
   \caption{Averaged spectrum of the six spectrum presented in this work, normalized at 0.60 $\mu$m. The positions of the absorption bands of CH$_4$ are represented by dotted lines.}
   \label{fig3}

\end{figure}

\begin{figure}
  \begin{center}
  \includegraphics[width=9.5cm]{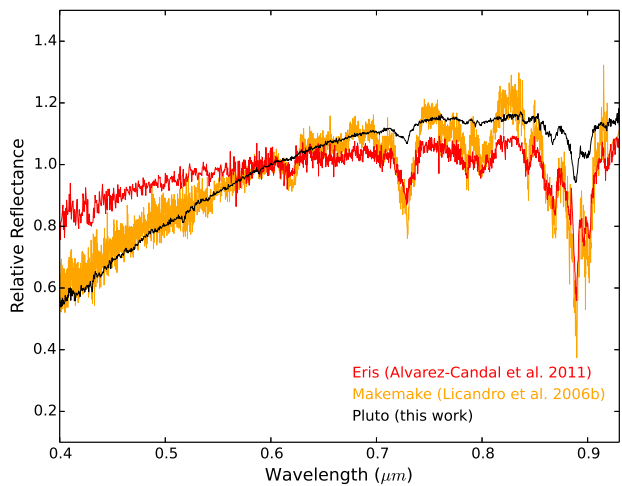}
 \end{center}
   \caption{Comparison of the spectra of Eris, Pluto and Makemake.  The three spectra are normalized at 0.60 $\mu$m. 
   All three objects show the absorption bands of methane ice. The band at 0.62 $\mu$m is more evident on Makemake and Eris than on Pluto. 
   Note the similarity between the slope of Pluto and Makemake that suggests the presence on both surfaces of a very similar coloring agent.}
   \label{fig4}

\end{figure}

The new spectra clearly show methane ice absorption bands at 0.73, 0.78 - 0.80, and 0.83 - 0.91 $\mu$m. 
In Fig. \ref{fig3}, we show the spectrum of Pluto obtained averaging the six spectra presented in this work, with the methane ice absorption bands evidenced as vertical dotted lines. 
All these bands have very low absorption coefficients, 5.0x10$^{-2}$, $\sim$1.3-1.4x10$^{-2}$, $\sim$1.0-8.0x10$^{-2}$ cm$^{-1}$, respectively, as measured in the laboratory at 40 K \citep{grund}. 
This average spectrum also shows an absorption around 0.62 $\mu$m. 
This band has been also observed in the spectra of other large TNOs dominated by methane ice, like Eris and Makemake \citep[][see Fig. \ref{fig4}]{Alvaro, licmak}, so its identification is not in question.  
However, the absorption coefficient of this band is unknown as it has not been measured in the laboratory.  
If we look at each particular reflectance obtained in 2014, some of them show it very clearly, in particularly the phase centered at 358.9$^{\circ}$. 

\begin{figure}
  \begin{center}
  \includegraphics[width=9.5cm]{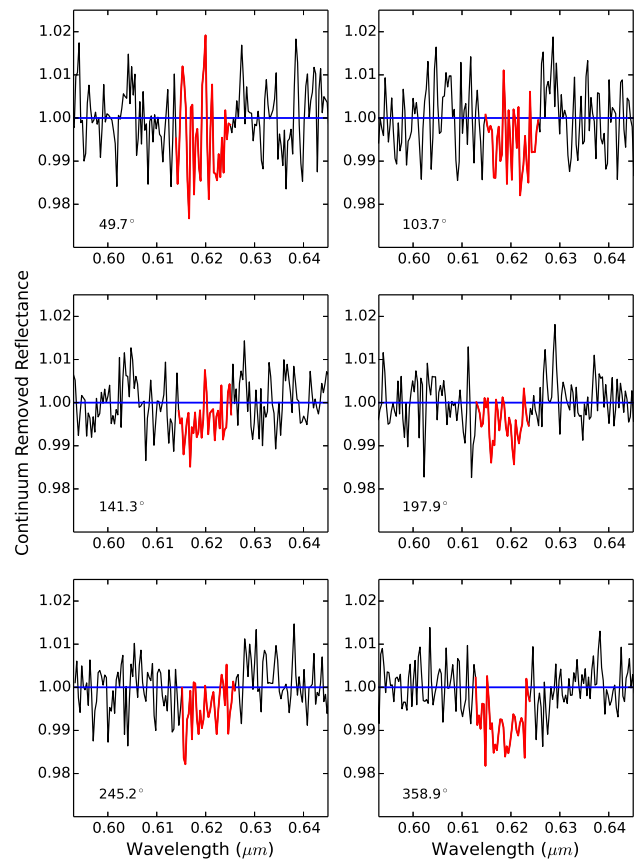}
 \end{center}
   \caption{Continuum-removed in the region of the spectra around the 0.62 $\mu$m absorption band of methane. In red the absorption bands, in blue the continuum.}
   \label{fig5}

\end{figure}

To confirm our detection, we compute the integrated area of the absorption band. For this computation, we first fit a straight line 
to the continuum on both sides of the band, between 0.593 – 0.615 $\mu$m and 0.624 – 0.653 $\mu$m, and divide the spectrum by this continuum (see Fig. \ref{fig5}). 
A first look at the plots in Fig. \ref{fig5} suggests a detection of the band, 
although in some of the phases (e.g. 49.7$^\circ$ or 103.7$^\circ$) it is not clear if the absorption is significant above the noise.

We then calculate the area of the absorption band after division by the continuum, integrating it between 0.615 and 0.624 $\mu$m. 
We propagate the errors, taking into account in each point the error associated with the relative reflectance and the error in the calculation of the fit of the continuum. 
Finally, we iterate the computation of the area, changing randomly the limits of the band, 
and we take the mean value and the standard deviation as the final values for the area and the associated error. 
Notice that for all the spectra (even those where the detection was suspicious) the integrated area of the band is a positive value. 
Some of the integrated areas, however, have large uncertainties. To confirm the detection and discard a serendipitous result, we do a second verification.

\begin{table}
 \caption{Integrated area in unit of 10$^{-4}$ of the absorption band at 0.62 $\mu$m and 
 test on the continuum between 0.55 and 0.57 $\mu$m measured as explained in the text. }
 \label{tab4}
 \centering                          
\begin{tabular}{l c c}        
\hline\hline 
\textbf{Longitude}& \textbf{Band at 0.62 $\mu$m}&\textbf{Cont. 0.55 – 0.57 $\mu$m}\\
\hline\\
 49.7$^{\circ}$ & 0.31 $\pm$ 0.30& 0.07 $\pm$ 0.70\\
 103.7$^{\circ}$& 0.39 $\pm$ 0.21& 0.51 $\pm$ 0.54\\
 141.3$^{\circ}$&0.29 $\pm$ 0.13 & 0.30 $\pm$ 0.35\\
 197.9$^{\circ}$& 0.33 $\pm$ 0.14& -0.11 $\pm$ 0.41\\
 245.2$^{\circ}$& 0.37 $\pm$ 0.14& 0.26 $\pm$ 0.46\\
 358.9$^{\circ}$& 0.66 $\pm$ 0.16& -0.57 $\pm$ 0.31\\
average spectrum& 0.35 $\pm$ 0.09& 0.13 $\pm$ 0.17\\
\hline                                   
\end{tabular}
\end{table}

For this test we repeat the procedure on a region of the continuum where no band is expected. We define this continuum in a region with the same number of points between 0.55 and 0.57 $\mu$m. 
What we could expect from this test is that we obtain both positive and negative values for the area, and that the absolute values of the areas would be smaller than the uncertainty. 
We show in Table \ref{tab4} the results of this test. Notice that for the calculation on the continuum, we obtained both positive and negative values, 
whereas for the absorption bands at 0.62 $\mu$m we obtained all positive values. 
After this test, we can consider the detection of the absorption bands at 0.62 $\mu$m as reliable, 
with the phase centered at $\sim$360$^\circ$ as the most clear detection.

\subsection{Band depth}
\label{depth}

To evaluate the globally averaged distribution of CH$_4$-ice around Pluto's surface, we computed the depth of the absorption bands as 
D = 1-R$_b$/R$_c$, where R$_b$ is the reflectance in the center of the band, and R$_c$ the reflectance of the continuum at the same wavelength. 
As several of the absorption bands of CH$_4$ in our spectra are a superposition of different absorption bands, 
it is not possible to measure the integrated area or the equivalent width of each single band. 
We calculated the band depth that can be isolated: 0.62, 0.73, 0.79, 0.80, 0.84, 0.87, 0.89 and 0.90 $\mu$m for our six spectra of Pluto. 
To evaluate the uncertainty in the calculation of the depth we propagated the errors, taking into account the uncertainties associated 
with the value of the relative reflectance in the center of the band and with the value of the continuum at the same wavelength. 
In Table \ref{tab5} we list the values obtained for the depths (in $\%$) of the six spectra of Pluto and the corresponding errors.

\begin{table*}
 \caption{Band depth ($\%$) and one-sigma errors for eight methane bands at each of the six longitudes observed.}
 \label{tab5}
 \centering                          
\begin{tabular}{l c c c c c c c c}        
\hline\hline 
\textbf{Longitude}& \textbf{0.62 $\mu$m}&\textbf{0.73 $\mu$m}& \textbf{0.79 $\mu$m}&\textbf{0.80 $\mu$m}&\textbf{0.84 $\mu$m}&\textbf{0.87 $\mu$m}&\textbf{0.89 $\mu$m}&\textbf{0.90 $\mu$m}\\
\hline\\
 49.7$^\circ$ & 1.16 $\pm$ 0.96& 4.97 $\pm$ 0.74& 2.57 $\pm$ 1.23&2.52 $\pm$ 1.10 & 1.49 $\pm$ 1.63& 7.07 $\pm$ 1.59& 17.35 $\pm$ 1.34& 10.19 $\pm$ 0.81\\
 103.7$^\circ$& 0.89 $\pm$ 0.74&4.27 $\pm$ 1.03 & 0.97 $\pm$ 0.58& 1.12$\pm$  0.68&1.54 $\pm$ 0.59 &5.69 $\pm$ 1.64 & 16.03 $\pm$ 1.40& 9.98$\pm$  0.55\\
 141.3$^\circ$& 0.50 $\pm$ 0.48&4.08 $\pm$ 0.90 & 1.33 $\pm$ 1.20&1.50 $\pm$ 1.04 & 1.59$\pm$  1.56&5.13 $\pm$ 1.98 & 15.77$\pm$  2.34& 9.85 $\pm$ 0.54\\
 197.9$^\circ$& 0.75 $\pm$ 0.41& 5.16 $\pm$ 0.46& 0.93$\pm$  0.57& 1.58 $\pm$ 0.53&1.05$\pm$  0.46 & 6.68 $\pm$ 2.64& 16.74$\pm$  1.17& 9.89 $\pm$ 1.44\\
 245.2$^\circ$& 0.89 $\pm$ 0.67& 4.99 $\pm$ 1.04& 1.77$\pm$  0.44&2.37 $\pm$ 0.71 & 1.90 $\pm$ 1.05&6.28 $\pm$ 1.87 & 17.33$\pm$  1.53& 11.01$\pm$  0.73\\
 358.9$^\circ$&1.13 $\pm$ 0.38 &6.29 $\pm$ 0.86 & 1.85 $\pm$ 0.86& 3.00 $\pm$ 0.61&1.75 $\pm$ 0.69 & 8.47 $\pm$ 1.49& 19.13 $\pm$ 1.41&10.41 $\pm$ 1.09 \\
\hline                                   
\end{tabular}
\end{table*}

\subsection{Lambert Coefficients}

The appearance of an intrinsically weak band like the one at 0.62 $\mu$m shown in Sec. \ref{depth} 
demonstrates a very long (several cm) path-length of sunlight in the nitrogen ice in which CH$_4$ is incorporated as a trace constituent. 
This, in turn, helps to set the size scale for the individual elements of polycrystalline solid N$_2$ characteristic of the surface of Pluto.

\begin{figure}
   \begin{center}
  \includegraphics[width=9.5cm]{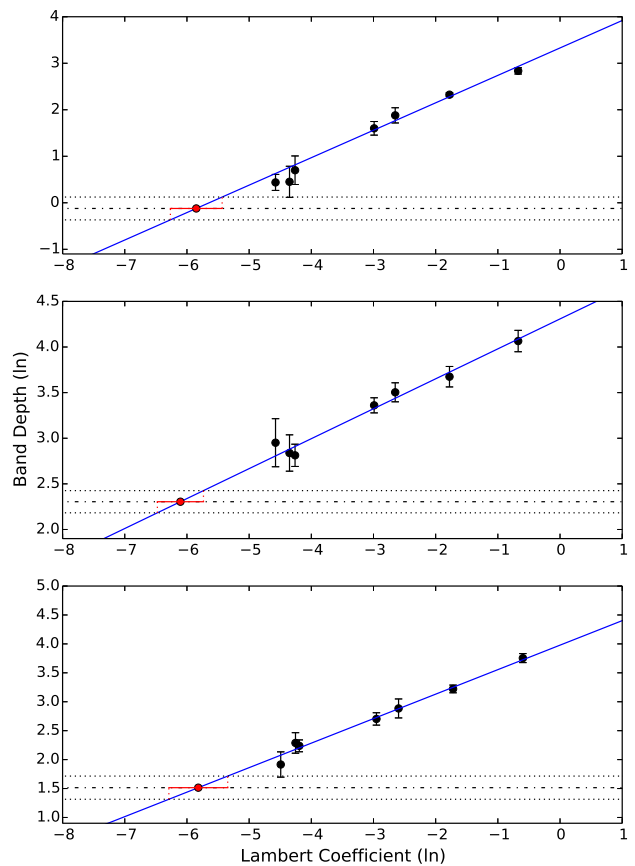}
 \end{center}
   \caption{Band depth (in natural logarithmic scale) vs Lambert coefficients (on natural logarithmic scale) for Pluto (top panel), Makemake (middle panel) and Eris (bottom panel).}
   \label{fig6}

\end{figure}

A great effort has been made to estimate the absorption coefficients of methane ice by means of laboratory experiments \citep[e.g.][]{grund}. 
However, these experiments do not typically cover wavelengths below 0.70 $\mu$m, 
as a large slab of methane is needed to produce that absorption. We will use this band observed in 
the spectrum of Pluto, Eris, and Makemake to estimate the absorption coefficient of CH$_4$ ice at 0.62 $\mu$m. 
For this purpose, we plotted on a  logarithmic scale the estimated band depth for the bands in the 0.73-0.90 $\mu$m range for the spectra of Pluto 
(for each band we used the average of the six values obtained for the six spectra) as a function of the corresponding Lambert coefficients at T = 40 K from \cite{grund}. 
Finally we fit a 1-degree polynomial (i.e. straight line) to the points with a least squares fit (see Fig. \ref{fig6} top panel).  

We do the same for the spectrum of Makemake from \cite{licmak} and for the spectrum of Eris from \cite{Alvaro}. 
In the case of Eris, we used the Lambert coefficients corresponding to T = 30 K, as this temperature is more similar to that estimated for the surface of this object (see Fig. \ref{fig6} bottom panel). 
In both cases we measured the band depth of methane ice in the spectrum of Makemake and Eris from the literature following the same procedure described in the previous section for Pluto. 

In all the plots in Fig. \ref{fig6}, the blue line is the least squares fit, the discontinued line corresponds to the measured band depth at 0.62 $\mu$m, with its corresponding error (dotted lines). 
The intersection between the line fitting the points and the measured band depth value for 0.62 $\mu$m band (red point) 
represents the extrapolated value of the Lambert coefficient for the 0.62 $\mu$m band, and its error (red line). 
In Table 6 we listed the band depths ($\%$), 
with the respective errors, and the value of the Lambert coefficient for each band, taken from the literature. 

\begin{table*}
  \caption{Band depths ($\%$), with respective errors, and Lambert coefficients \citep{grund} used for the extrapolation of the Lambert coefficient at 0.62 $\mu$m. 
  The values of the coefficient in bold format have been estimated in this work. 
  *average of the values computed for the six spectra shown in Table \ref{tab5}; 
  **band depth measured for the spectrum of Makemake from \cite{licmak} and for the spectrum of Eris from \cite{Alvaro}.}
 \label{tab6}
 \centering                          
\begin{tabular}{l c c c c c c c c}        
\hline\hline 
\textbf{Object}& \textbf{0.62 $\mu$m}&\textbf{0.73 $\mu$m}& \textbf{0.79 $\mu$m}&\textbf{0.80 $\mu$m}&\textbf{0.84 $\mu$m}&\textbf{0.87 $\mu$m}&\textbf{0.89 $\mu$m}&\textbf{0.90 $\mu$m}\\
\hline\\
 Pluto *& 0.9 $\pm$ 0.3& 5.0 $\pm$ 0.8&1.6 $\pm$ 0.6 &2.0 $\pm$ 0.7 & 1.6 $\pm$ 0.3& 6.6 $\pm$ 1.2& 17.1$\pm$ 1.2&10.2 $\pm$0.4 \\
 Makemake**& 10.0 $\pm$ 1.3&28.8 $\pm$  2.5 &17.1  $\pm$ 3.8 &16.7 $\pm$ 2.2 &19.1 $\pm$ 5.8 & 33.2 $\pm$ 3.7&58.3 $\pm$ 7.3 & 39.4 $\pm$ 4.7\\
 Lambert Coef. (40 K)&\textbf{2.23-2.88x10$^{-3}$} &5.03x10$^{-2}$ & 1.29x10$^{-2}$& 1.41x10$^{-2}$& 1.03x10$^{-2}$&7.06x10$^{-2}$ &5.10x10$^{-1}$ &1.69x10$^{-1}$ \\
\hline
&&&&&&&&\\
 Eris**&4.6 $\pm$  1.0 & 14.9 $\pm$ 1.7&9.9 $\pm$  1.9 & 9.4 $\pm$ 1.0&6.8 $\pm$  1.7 &17.9 $\pm$ 3.2 &42.7 $\pm$ 3.5 &25.1 $\pm$ 1.7 \\
 Lambert Coef. (30 K)&\textbf{2.97x10$^{-3}$} &5.22x10$^{-2}$ & 1.42x10$^{-2}$& 1.51x10$^{-2}$& 1.12x10$^{-2}$& 7.45x10$^{-2}$& 5.50x10$^{-1}$& 1.79x10$^{-1}$\\
\hline                                   
\end{tabular}
\end{table*}

The value of the extrapolated Lambert coefficients at 0.62 $\mu$m for Pluto is 2.88x10$^{-3}$ $\pm$ 1.23x10$^{-3}$ cm$^{-1}$, while the value obtained for Makemake is 2.23x10$^{-3}$ $\pm$ 0.85x10$^{-3}$ cm$^{-1}$. 
The two values are in agreement within the errors as expected, since both objects have similar estimated surface temperature.

The value obtained for Eris is 2.97x10$^{-3}$ $\pm$ 1.46x10$^{-3}$ cm$^{-1}$. The difference due to the temperature is not significant within the error in the calculation.

\subsection{Shifts of the center of the bands}

The centers of the absorption bands of methane ice provide information about the mixing ratio of methane and nitrogen \citep{quirico}. 
When methane is diluted in nitrogen ice in small proportions, the centers of the methane bands are shifted towards bluer wavelengths \citep{quirico}, 
and the shift is larger as the content of nitrogen increases \citep{brunetto2008}. 
Some works \citep{quirico,quirico1999,doute,merlin2010} discuss the existence of regions formed of pure CH$_4$ segregated from the N$_2$:CH$_4$ solution. 
This effect was first reported for Pluto by \cite{owen}. 
However, this is in conflict with both alternative models for explaining Pluto's atmosphere. According to \cite{trafton2015}, and references therein,
the surface of Pluto would instead be covered by a solid solution of these ices, N$_2$:CH$_4$, with different mixing ratios. 
Moreover, two dilution regimes, one saturated in CH$_4$ and the other saturated in N$_2$ would coexist on the surface of Pluto.
The dilution regime saturated in CH$_4$ would be the responsible of the relatively unshifted part of the Pluto's spectra, 
previously attributed to pure CH$_4$. 

We measured the shifts of the center of the bands by comparing our spectra with theoretical relative reflectances of pure methane ice. 
The synthetic relative reflectances were calculated using a Shkuratov model \citep{shku} and the optical constants of methane for T = 40 K from \cite{grund}. 

For comparison, we obtained a suit of relative reflectances considering different grain sizes, then we shifted each model in wavelength until we found the best comparison.
We used a chi-square test to choose the best combination of grain size and shift. 
To evaluate the error associated with the measurement, we repeated the process several times, excluding randomly some points of our spectra and changing the limits of the bands. 
The obtained shifts are listed in Table \ref{tab7}. In some cases, due to the low signal to noise 
or to the presence of some artifacts that appear during the reduction process and change the shape of the band, it has been impossible to measure the band shift.

\begin{table*}
  \caption{Shifts (\AA) of the center of the absorption band of CH$_4$.}
 \label{tab7}
 \centering                          
\begin{tabular}{l c c c c c c c c}        
\hline\hline 
\textbf{Longitude}& \textbf{0.6190 $\mu$m}&\textbf{0.7296 $\mu$m}& \textbf{0.7862 $\mu$m}&\textbf{0.7993 $\mu$m}&\textbf{0.8442 $\mu$m}&\textbf{0.87 $\mu$m}&\textbf{0.8897 $\mu$m}&\textbf{0.9019 $\mu$m}\\
\hline\\
  49.7$^\circ$& -& -7.4 $\pm$ 4.6 & -         & -          & -         & -          & -7.6$\pm$  1.5 & -\\
 103.7$^\circ$& -& -15.6 $\pm$ 2.7& -         & -          & -9.3 $\pm$ 1.7& -13.9 $\pm$ 2.2& -13.9 $\pm$ 1.8& -\\
 141.3$^\circ$& -& -          & -         & -          & -         & -          & -          & -\\
 197.9$^\circ$& -&-12.1 $\pm$ 1.4 & -         & -14.8$\pm$  2.5& -6.9 $\pm$ 2.7& -10.8 $\pm$ 6.3& -15.7 $\pm$ 3.4& -\\
 245.2$^\circ$& -& -5.7 $\pm$ 1.8 & -         & -          & -         & -5.5 $\pm$ 2.6 & -          & -\\
 358.9$^\circ$& -& -4.7 $\pm$ 2.7 & -7.1 $\pm$ 2.0& -4.0 $\pm$ 3.0 & -         & -3.1 $\pm$ 2.6 & -8.8 $\pm$ 2.3 & -\\
\hline                                   
\end{tabular}
\end{table*}

\section{Discussion}

Previous works have shown that Pluto presents a surface with a zonal (longitudinal) heterogeneity (\citealt{grundy1996}; \citealt{Buie_a, Buie_b}; G13). 
This heterogeneity is clearly seen in processed images of Pluto from the HST \citep{Buie_b} 
and in the recently released images of the \textit{New Horizons} mission (www.nasa.gov/mission$\_$pages/newhorizons). 
The heterogeneity is associated with an uneven distribution of the ices (CH$_4$, N$_2$, CO) and the complex organics on the surface. 

G13 report long time trends in near-infrared spectra obtained between 2000 and 2012 from the analysis of the absorption bands of CH$_4$, CO, and N$_2$ present in the spectra. 
They find that the absorptions of CO and N$_2$ are concentrated on Pluto's anti-Charon hemisphere ($\sim$180$^\circ$), unlike absorptions of less volatile CH$_4$ ice 
that are offset by roughly 90 degrees from the longitude of maximum CO and N$_2$ absorption. In the time scale of their observations, 
they find variations in the depth and in the amplitude of the diurnal variation of the absorption bands of CH$_4$, CO and N$_2$. 
They conclude that the geometric distribution of these ices (in particular CO and N$_2$) on the surface of Pluto can explain these secular changes, 
although seasonal volatile transport could be at least partly responsible. 

Although visible wavelength spectroscopy does not enable the study of the N$_2$ and CO concentrations, 
we can discuss the longitudinal variation within our set of observations and the secular variation when comparing with previous observations, 
of the distribution of methane-ice and of the dark-red organic material. 

Also, from the study of the center of the CH$_4$ bands, we can extract information on the relative amount of CH$_4$ diluted in N$_2$. 

With respect to the methane, G13 find that the stronger CH$_4$ absorption bands have been getting deeper, while the amplitude of their diurnal variation is diminishing, 
consistent with additional CH$_4$ absorption at high northern latitudes rotating into view as the sub-Earth latitude moves north.  
They conclude that the case for a volatile transport contribution to the secular evolution looks strongest for CH$_4$ ice, despite it being the least volatile of the three ices.

\subsection{Longitudinal variations}

\begin{figure*}
    \begin{center}
  \includegraphics[width=15cm]{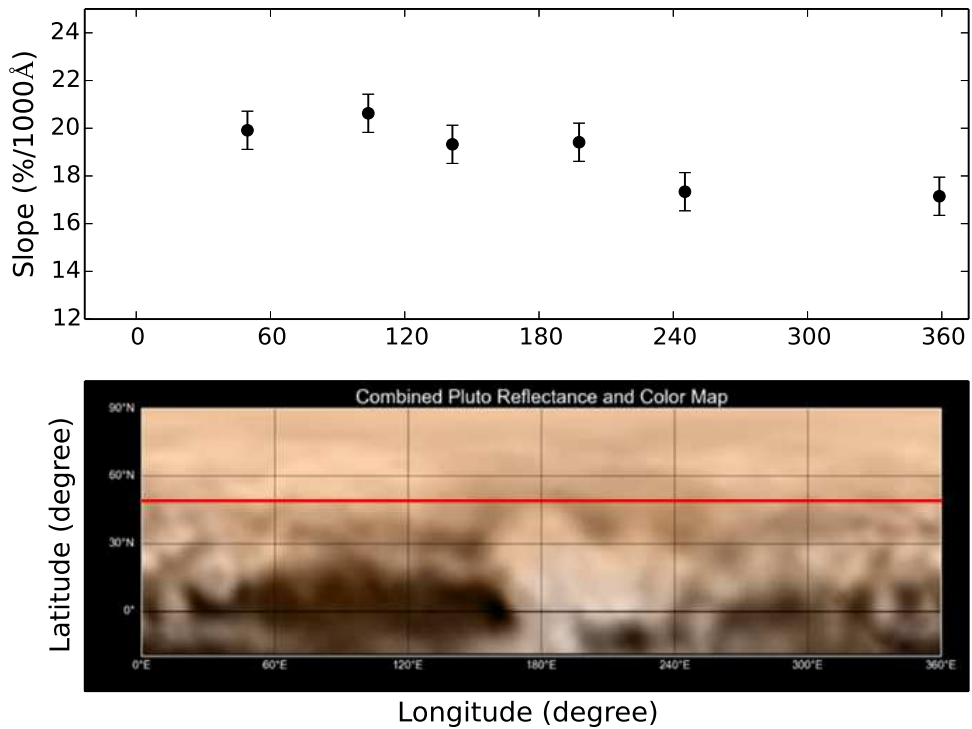}
 \end{center}
   \caption{Spectral slope ($\%/1000$\ \AA), calculated between 0.40 and 0.75 $\mu$m for spectra of Pluto normalized at 0.60 $\mu$m vs longitudes (top panel), and map of part 
   of the surface of Pluto result of combination of images taken by the Long Range Reconnaissance Imager (LORRI) on \textit{New Horizons}, and combined 
   with the lower-resolution color data from the instrument Ralph on the spacecraft  (bottom panel) (www.nasa.gov/mission$\_$pages/newhorizons). 
   The scale of the horizontal axis is the same for the top and bottom panel. 
   The red line indicate the subEarth latitude at which our data have been obtained ($\sim$51$^\circ$).}
   \label{fig7}

\end{figure*}

We note from Fig. \ref{fig2} that spectra at different longitudes have different albedos as well as different overall shapes across the full spectral range, demonstrating the longitudinal heterogeneity of the surface of Pluto. 
To show this in a more clear way, in Fig. \ref{fig7} we plotted the variation with 
longitude of the spectral slope value obtained from our spectra, and compared it with a map of part of the surface of Pluto (mostly the northern hemisphere 
which is also the hemisphere currently observed from Earth) obtained by the \textit{New Horizon} mission (www.nasa.gov/mission$\_$pages/newhorizons).  Notice that highest 
(redder) value of the slope (at longitude $\sim$100$^\circ$) is centered on the hemisphere that, according to the new maps of Pluto, contains the large dark region, 
whereas lower values correspond with a zone in average brighter ($\sim$240$^\circ$). This is compatible with a higher abundance of solid organic material 
(redder in color and lower in albedo than the icy species observed in Pluto) in the darker region and a higher abundance of ices in the brighter ones. 

\begin{figure}
    \begin{center}
  \includegraphics[width=9.5cm]{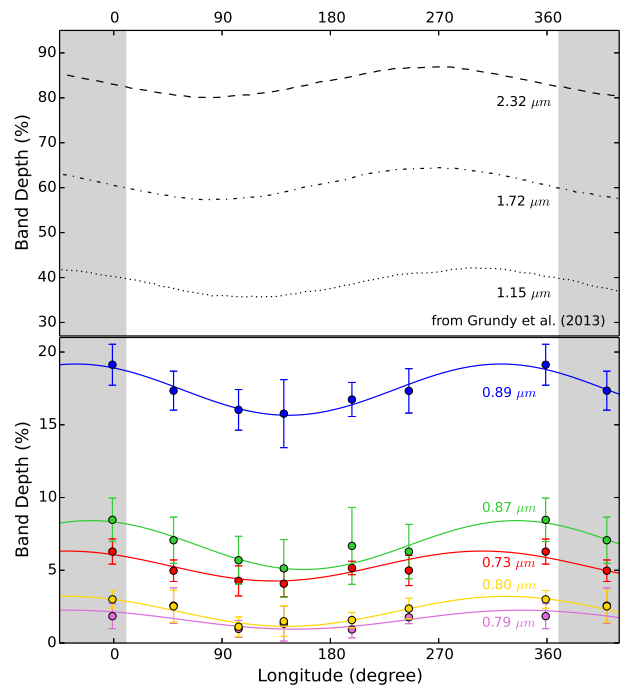}
 \end{center}
   \caption{Bottom Panel: variation of the band depth (in $\%$) with longitude for the five stronger CH$_4$ absorption bands measured from the spectra of Pluto presented in this work. 
   Top panel: variation of the depth of the absorption bands at 1.14, 1.72 and 2.32 $\mu$m from G13. 
   The points in the gray shadow represent replicate values (at longitude $\pm$ 360$^\circ$) to better visualize the variation of the band depth with the longitude.}
   \label{fig8}

\end{figure}

In Fig. \ref{fig8} (bottom panel), we show the variation of the band depth of the five stronger CH$_4$ absorption bands measured in our spectra. 
A sinusoidal trend is clearly visible with a minimum 
at a longitude between 140$^\circ$-160$^\circ$ and a maximum at a longitude near 360$^\circ$. This phase (360$^\circ$) is also the one with 
the clearest detection of the 0.62-$\mu$m band. The location of this CH$_4$ maximum 
is in agreement with the trend found by G13 (maximum CH$_4$ between 260$^\circ$ and 320$^\circ$), but appears a bit shifted to the East. 
In particular, G13 finds that there are subtle differences in the longitude of the maximum of the absorption for different bands of CH$_4$, with 
the weaker absorption bands having the minimum and maximum at longitudes more to the east than the stronger CH$_4$ absorption bands (see Fig. 4 of G13). 
They interpreted this difference between stronger and weaker bands as due to the presence of different terrain types, 
with terrains with larger particles, or with greater CH$_4$ enrichment, that produce the weaker bands more prevalent on the Charon-facing hemisphere. 
Fig. \ref{fig8} shows how this trend continues for the bands in the visible that are even weaker than in the NIR, 
with the shallowest bands (0.80 and 0.79 $\mu$m) having a minimum at $\sim$180$^\circ$ and a maximum at $\sim$360$^\circ$. 
The largest abundance of CH$_4$, or the largest particles, can be found in the hemisphere centered at $\sim$0$^\circ$, 
and is not coupled either with the darkest (probably higher organic abundance) or with the brightest areas (probably higher volatile component). 

\begin{figure}
    \begin{center}
  \includegraphics[width=9.5cm]{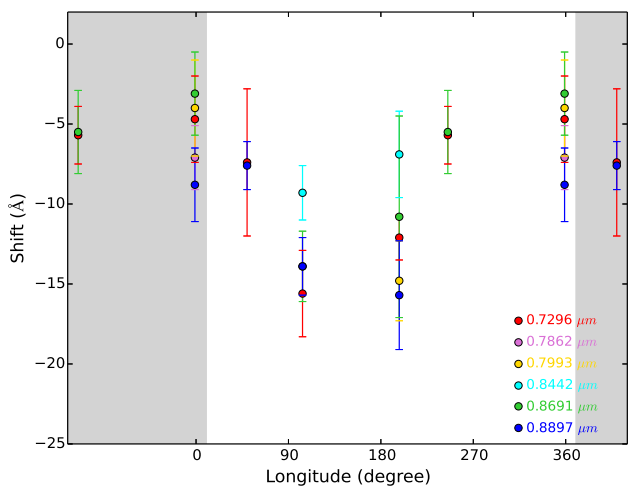}
 \end{center}
   \caption{Measured band shifts vs. Pluto's longitude. 
   Notice that at longitudes between 100$^\circ$ and 200$^\circ$ the shifts are larger, thus the CH$_4$:N$_2$ is smaller, 
   which is compatible with a lower abundance of CH$_4$ suggested by the smaller band depths shown in Fig. \ref{fig8}.}
   \label{shift}

\end{figure}

With respect to the degree of dilution of CH$_4$ in N$_2$, G13 found that all the CH$_4$ bands in the infrared range show a blue shift (indicative of dilution), 
with a minimum blue shift corresponding to the Charon-facing hemisphere, and a maximum blue shift corresponding to the anti-Charon hemisphere. 
That is exactly what Fig. \ref{shift} shows, with a maximum absolute value of the shifts in the anti-Charon hemisphere. 
The minimum of the shifts correspond with a region where the abundance of methane is higher (the Charon-facing hemisphere). 
This could indicate a higher degree of saturation of CH$_4$ in this hemisphere \citep{tegler2012, Grundy2013, trafton2015}. 
Our measurements in the visible show the same trend that the estimates from the NIR observations, 
apparently indicating that the degree of dilution is uniform over different path-lengths of the light.

\subsection{Secular variations}

\begin{figure}
    \begin{center}
  \includegraphics[width=9.5cm]{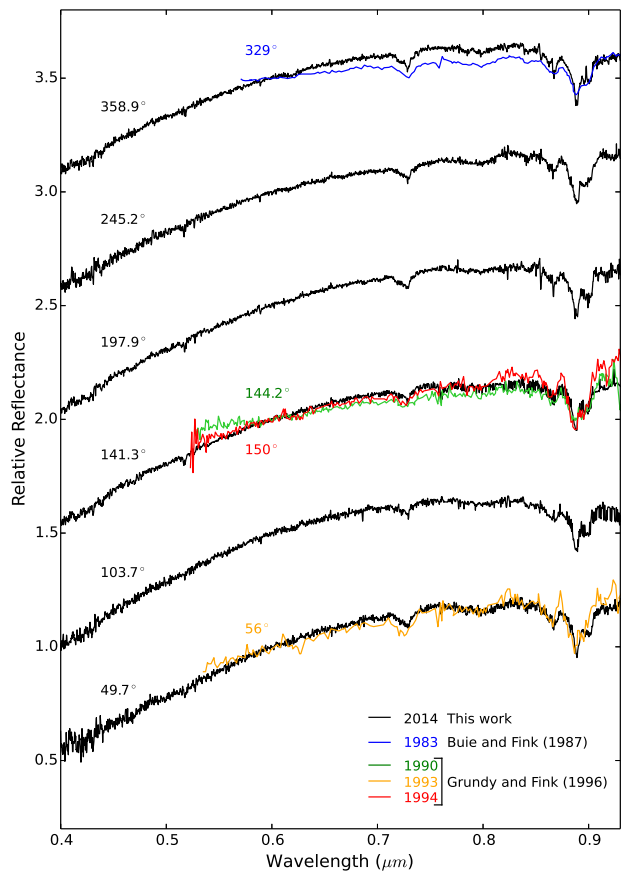}
 \end{center}
   \caption{Secular variation on Pluto reflectance. 
   Pluto-Charon spectra from this work shown with data at similar spectral resolution and signal precision from \cite{buie1987} and \cite{grundy1996} at similar latitudes. 
   All the spectra are normalized at 0.60 $\mu$m and shifted by 0.5 in relative reflectance for clarity.}
   \label{compgrund}

\end{figure}

In order to compare our observations with earlier spectral data, in Fig. \ref{compgrund} we show all our spectra, 
with spectra from \cite{buie1987} and \cite{grundy1996} overplotted on spectra obtained at similar longitudes (but more northern latitudes).  
For example, we compare our spectrum taken at a longitude of 49.7$^\circ$ (subEarth latitude 51$^\circ$), 
with Grundy and Fink's data for 1993 (longitude 56$^\circ$, subEarth latitude $\sim$11)$^\circ$, 
and our spectrum at 358.9$^\circ$ longitude (subEarth latitude 51$^\circ$) with 1983 spectrum from \cite{buie1987} (longitude 329$^\circ$, subEarth latitude $\sim$ -10$^\circ$). 
Differences in the new spectra compared to earlier data can be a result of the different latitude aspect, temporal changes, or a combination of both.
From the comparision of the slopes in the visible (see Table \ref{tab3}), 
we can see that the spectra obtained in 2014 are all redder than previous spectra. 
This could suggest a stronger presence of complex organics on Pluto's surface with respect to previous epochs.

\begin{figure}
    \begin{center}
  \includegraphics[width=9.5cm]{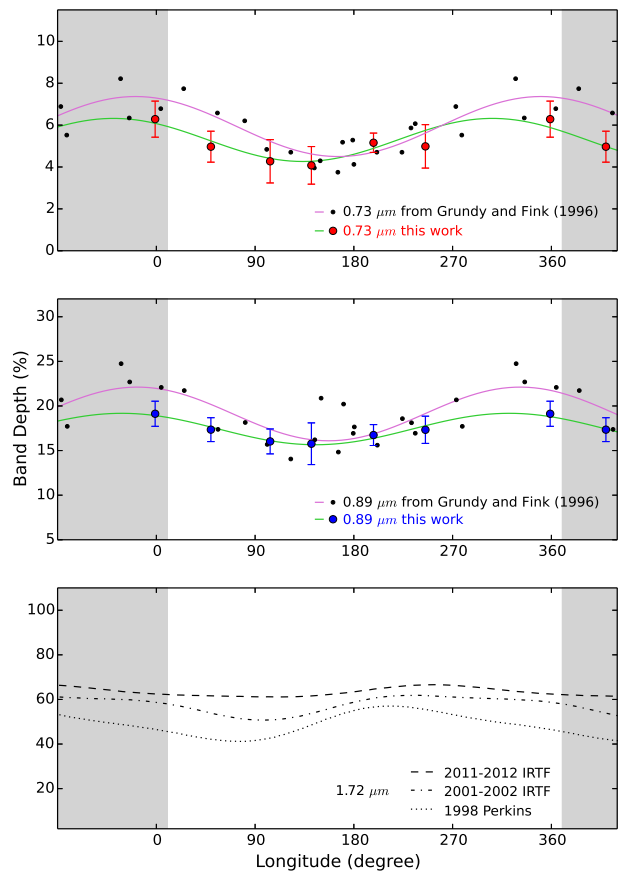}
 \end{center}
   \caption{Secular trend of the methane absorption bands: variation of the absorption band depth for the band at 0.73 $\mu$m (top panel) and 0.89 $\mu$m (middle panel) 
   compared with the secular variation for the band at 1.72 $\mu$m (bottom panel) from G13. 
   Lines in green represent a sinusoidal fit of band depths from this work; 
   lines in red represent a sinusoidal fit of band depths from \cite{grundy1996}. 
   The points in the gray shadow represent replicate values (at longitude 360$^\circ$) to better visualize the variation of the band depth with the longitude.}
   \label{secdepth}

\end{figure}

G13 reported a difference in the secular evolution between stronger and weaker methane absorption bands observed in the near-infrared. 
They found that the stronger bands increase in strength with years, while the weaker bands in their sample do not. 
If this trend extends to the visible wavelength we would see a secular decrease of the depth of the bands. 
Indeed, from Fig.\ref{secdepth}, we can note that the depths of the bands at 0.73 and 0.89 $\mu$m measured in our spectra 
are lower than those reported in \cite{grundy1996}, as we can expected for weak bands.

G13 also found that the amplitude of the pattern of the longitudinal variation decreases with time. 
This is consistent with the change over the years of the sub-Earth latitude of Pluto, 
with an increasing contribution to the observed spectrum of northern polar ices  
over the contributions from equatorial and southern latitudes, which show a much greater albedo contrast with longitude. 
In Fig. \ref{secdepth} (top and middle panel) we plotted the measured band depth of the weak bands at 0.73 and 0.89 $\mu$m, 
along with the depth of the same bands from \cite{grundy1996}, and a sinusoidal fit for each data set. 
In the bottom panel we show, for comparison purposes, the secular variation of the strong band at 1.72 $\mu$m from G13. 
The depths and the amplitude of the variations measured by us 
are slightly lower than those of 1980 and 1994, following the same trend of the near-infrared bands described in G13. 

In summary, the analysis of the secular variation of the slope in the visible, shows a reddening of Pluto's surface from 1983 to 2014, 
suggesting a higher degree of processing of the surface. 
On the other hand, the analysis of the secular variation of the methane ice bands in the visible is consistent with the results from the NIR, 
and indicates a decrease of the contrasts in band depth along Pluto's surface.
The clear correspondence of these changes with the pattern seen in the albedo maps of Pluto obtained with HST 
and \textit{New Horizons} supports the contention that they are probably dominated by the sub-Earth latitudinal change rather than by seasonal changes.

\subsection{Comparison with Makemake and Eris}

To compare the spectral properties of Pluto, Makemake and Eris, 
in Sec. \ref{res_an} we calculate for both the same parameters computed for Pluto's spectra: slope and band depths. 

The spectral slope of Pluto calculated between 0.57 and 0.70 $\mu$m is identical within the errors to that of Makemake from \cite{licmak}  
and three times larger than the spectral slope of Eris from \cite{Alvaro} (see Table \ref{tab3} and Fig. \ref{fig4}). This suggests 
that the surface of Makemake contains a similar amount of solid complex organics (the coloring agent) as Pluto (averaging over the entire surface), 
while Eris' surface has a smaller amount of organics than both Pluto and Makemake, as previously suggested \citep[e.g. by][and references therein]{lor_make}. 
This shallower spectral slope is consistent with refreshing of the surface of Eris with volatile ices (N$_2$, CO, CH$_4$) deposited 
by the collapse of its atmosphere when this TNO was farther from the Sun, as suggested by \cite{liceris} and \cite{Alvaro}. 

On the other hand, the methane bands are consistently deeper in the Makemake spectrum than in the Eris spectrum, 
and in the Pluto spectrum they are even weaker (see Table \ref{tab6} and Fig. \ref{fig4}). 
This is indicative that methane is more abundant on the surface of Makemake than on that of Eris, 
and on the surface of Eris than on that of Pluto, as previously reported by \cite{licmak}. 
The two major known mechanisms that destroy methane ice on the surface of Pluto are  
transformation into red complex organics under energetic irradiation and sublimation of methane into the atmosphere. 
As the overall colors of Pluto and Makemake are so similar, we infer that the more plausible explanation 
for the difference in the abundance of methane on Pluto and Makemake is the existence of a bounded atmosphere around Pluto.

\subsection{Other applications}
\begin{figure}
    \begin{center}
  \includegraphics[width=9.5cm]{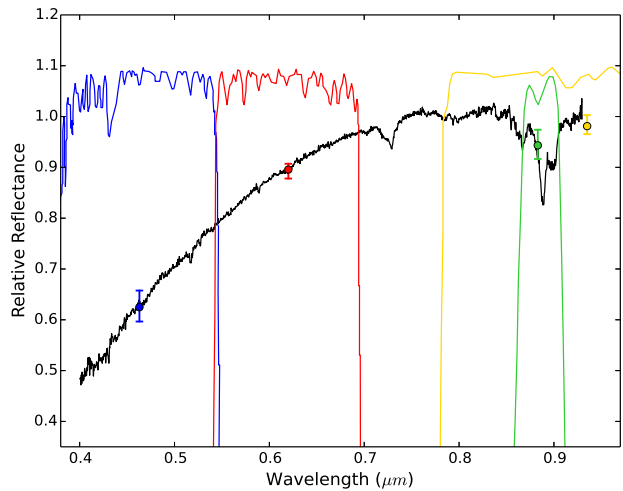}
 \end{center}
   \caption{Longitudinal variation inferred for MVICS observations. The average spectrum from the 2014 observations presented in this work is presented together 
   with the filter transmission bands of the New Horizons Multispectral Visible Imaging Camera (MVIC) \citep{reuter2008}. 
   The colored dots show the convolution of these transmissions 
   with the average Pluto spectrum. The vertical bar shows the variation among the six reflectances (maximum and minimum value). 
   The value at the NIR band is a lower limit as our spectra do not cover all the band-pass of that filter.}
   \label{mvic}

\end{figure}

The spectra presented in this paper provide an independent calibration of the MVIC (Multispectral Visible Imaging Camera) instrument aboard the \textit{New Horizons} spacecraft, 
and allow the synthesis of a “green channel” that enable the preparation of false-color images of Pluto with the data from the spacecraft encounter.  
Fig. \ref{mvic} shows an average spectrum of Pluto from the present investigation with the transmission curves of the MVIC filters superimposed. 
Overplotted we show the photometric appearance of this spectrum as seen through MVIC filters. 
The vertical lines indicate the maximum and minimum value of these “convolved photometric” points 
from our set of spectra and give an indication of how the variation showed by our data would be detected by MVIC. 

These new spectra provide the desired color information for modeling efforts that will use optical constants for non-ice materials presently being synthesized 
in the laboratory by the UV and electron-irradiation of a Pluto mixture of ices known from near-IR spectroscopy (N$_2$:CH$_4$:CO, 100:1:1). 
The chemical analysis of the refractory residue from a UV-irradiated Pluto ice mixture has been published by \cite{materese2014}, 
and results from the electron-irradiated ices are presented in Materese et al. (submitted).

\section{Conclusions}

Low-resolution visible spectra of Pluto obtained at six different planetocentric longitudes between May 13 and  July 17,  2014 were obtained using the ACAM 
imager/spectrograph of the 4.2 m William Herschel Telescope (Observatorio del Roque de los Muchachos, La Palma, Spain). 
These data cover a whole rotation of Pluto with a step of $\sim$50$^\circ$, and a gap of $\sim$300$^\circ$.

The new spectra clearly show methane ice absorption bands at 0.73, 0.78 - 0.80 $\mu$m, and 0.83 - 0.91 $\mu$m also seen in previously reported spectra of Pluto \citep{grundy1996}. 
We report the detection of the spectra of the methane-ice band equivalent of the 0.62-$\mu$m CH$_4$ gas absorption band in Pluto spectrum. 
This band was previously reported in the spectra of TNOs Eris and Makemake \citep{Alvaro, licmak}. 
The measurement of the depth of the absorption band of methane ice at 0.62 $\mu$m in the new six spectra of Pluto allowed us 
to estimate the Lambert coefficient at this wavelength at T = 40$^\circ$ K (not measured in the laboratory yet) as 2.88x10$^{-3}$ $\pm$ 1.23x10$^{-3}$ cm$^{-1}$. 
From the Makemake spectrum, we obtained a similar value within errors, 2.23x10$^{-3}$ $\pm$ 0.85x10$^{-3}$ cm$^{-1}$. 
From the spectrum of Eris, we obtained a Lambert coefficient of 2.97x10$^{-3}$ $\pm$ 1.46x10$^{-3}$ cm$^{-1}$, in this case for a T = 30$^\circ$ K. 

The spectral properties of Pluto, compared to those of Makemake and Eris, suggest that Pluto's surface 
holds a similar amount of solid complex organics (the coloring agent) as Makemake, 
while Eris' surface has a lesser amount of organics than them both, 
as previously suggested \citep[e.g.][]{licmak, Alvaro, lor_make}. 
On the other hand, methane is more abundant on the surface of Makemake than on that of Eris, 
and on the surface of Eris than on that of Pluto as previously reported by \cite{licmak}. 

The longitudinal variation of the slope in the visible is in phase 
with the distribution of the dark and bright material as seen in the albedo maps of Pluto from HST and \textit{New Horizons}, 
with the reddest slopes in the hemisphere centered at 90$^\circ$ and the less red material in the hemisphere centered at 270$^\circ$. 

The longitudinal and secular variations of the methane bands observed in the visible wavelengths are consistent with the behavior of the weak absorption bands in the near-infrared region reported by G13. 
The maximum absorption due to methane ice is located at the phase centered at 360$^\circ$ (Charon-facing hemisphere) and the minimum at $\sim$180$^\circ$ (anti-Charon-facing hemisphere). 
The curve is shifted from the curves of the stronger methane ice bands in the NIR. 
The depths and the amplitude of the diurnal variations of the methane ice in the spectra presented here (54$^\circ$ N) are slightly lower than those of 1980 and 1994 \citep{grundy1996}, following the same trend of the near-infrared bands described in G13. 
This trend indicates that our observations capture regions with a lower contrast and is in agreement with the contrast in the maps of Pluto with more dark and bright material lying in the equatorial latitudes. 

All of the detected CH$_4$ bands are blue shifted, indicating a certain degree of dilution of the CH$_4$ in other ice, probably N$_2$. We observe a longitudinal variation 
with a maximum shift located at longitudes between 100$^\circ$ – 200$^\circ$ and a minimum shift around 360$^\circ$ longitude. The fact that the shift is minimum 
in the regions where the abundance of methane is higher could be indicative of a dilution of CH$_4$:N$_2$ more saturated in CH$_4$. 
 
\begin{acknowledgements}
 This paper is based on data from override observations made with the William Herschel Telescope, under program 055-WHT11/14A, operated on the island of La Palma 
 by the Isaac Newton Group in the Spanish Observatorio del Roque de los Muchachos of the Instituto de Astrof\'isica de Canarias.  
 DPC,  WMG, and RPB were supported in part by NASA’s New Horizons mission. 
 JL acknowledges support from the project ESP2013-47816-C4-2-P (MINECO, Spanish Ministry of Economy and Competitiveness).
\end{acknowledgements}

\bibliographystyle{aa} 
\bibliography{pluto} 

\end{document}